\documentclass[twocolumn,3p,preprint]{elsarticle}
\journal{Physics Letters B}
\usepackage[numbers]{natbib}
\RequirePackage{fancyhdr}
\RequirePackage{graphicx}
\usepackage{placeins}
\RequirePackage{adjustbox}
\usepackage{rotating}
\usepackage{multirow}
\usepackage{geometry}
\usepackage{ragged2e}
\usepackage{color}
\usepackage{array}
\usepackage{float}
\usepackage{units}
\RequirePackage{amssymb}
\RequirePackage{epstopdf}
\RequirePackage{amsmath, amsfonts}

\usepackage{graphicx}
\usepackage{dcolumn}
\usepackage{bm}
\usepackage{float}
\usepackage{xcolor}
\usepackage{hyperref}
\usepackage{tabularx}
\usepackage{multirow}

\biboptions{sort&compress}
\bibpunct{[}{]}{,}{n}{,}{,}

\begin{document}
\begin{frontmatter}

\title{Damping of the isovector giant dipole resonance in $^{40,48}$Ca}

\author[Wits]{J.~Carter}
\author[Wits,iTL]{L.M.~Donaldson}
\author[RCNP]{H.~Fujita}
\author[RCNP]{Y.~Fujita}
\author[Wits,Wits2]{M.~Jingo}
\author[Wits,Bots]{C.O.~Kureba}
\author[Wits,KUL]{M.B.~Latif}
\author[WMU]{E.~Litvinova}
\author[iTL]{F.~Nemulodi}
\author[TUDarm]{P.~von~Neumann-Cosel\corref{cor1}}\ead{vnc@ikp.tu-darmstadt.de}
\author[iTL]{R.~Neveling}
\author[RIP]{P.~Papakonstantinou}
\author[US]{P.~Papka}
\author[Wits,iTL]{L.~Pellegri}
\author[TUDarm]{V.Yu.~Ponomarev}
\author[TUDarm]{A.~Richter}
\author[TUDarm]{R.~Roth}
\author[Wits]{E.~Sideras-Haddad}
\author[iTL]{F.D.~Smit}
\author[iTL,US]{J.A.~Swartz}
\author[RCNP]{A.~Tamii}
\author[TUDarm]{R.~Trippel}
\author[Wits]{I.T.~Usman}
\author[AS]{H.~Wibowo}

\address[Wits]{School of Physics, University of the Witwatersrand, Johannesburg 2050, South Africa}
\address[iTL]{iThemba Laboratory for Accelerator Based Sciences, Somerset West 7129, South Africa}
\address[RCNP]{Research Center for Nuclear Physics, Osaka University, Ibaraki, Osaka 567-0047, Japan}
\address[Wits2]{School of Clinical Medicine, University of the Witwatersrand, Johannesburg 2000, South Africa}
\address[Bots]{Department of Physics and Astronomy, Botswana International University of Science and Technology, Palapye, Botswana}
\address[KUL]{KU Leuven, Instituut voor Kern- en Stralingsfysica, B-3001 Leuven, Belgium}
\address[WMU]{Department of Physics, Western Michigan University, Kalamazoo MI 49008-5252,USA}
\address[TUDarm]{Institut f\"ur Kernphysik, Technische Universit\"at Darmstadt, 64289 Darmstadt, Germany}
\address[RIP]{Rare Isotope Science Project, Institute for Basic Science, Daejeon 34000, South Korea}
\address[US]{Department of Physics, University of Stellenbosch, Matieland 7602, South Africa}
\address[AS]{Institute of Physics, Academia Sinica, Taipei, 11529, Taiwan}

\cortext[cor1]{Corresponding author}

\begin{abstract}

The fine structure of the IsoVector Giant Dipole Resonance (IVGDR) in the doubly-magic nuclei $^{40,48}$Ca observed in inelastic proton scattering experiments under $0^\circ$ is used to investigate the role of different mechanisms contributing to the IVGDR decay width.
Characteristic energy scales are extracted from the fine structure by means of wavelet analysis. 
The experimental scales are compared to different theoretical approaches allowing for the inclusion of complex configurations beyond the mean-field level.
Calculations are performed in the framework of RPA and beyond-RPA in a relativistic approach based on an effective meson-exchange interaction, with the UCOM effective interaction and, for the first time, with realistic two- plus three-nucleon interactions from chiral effective field theory employing the in-medium similarity renormalization group. 
All models highlight the role of Landau fragmentation for the damping of the IVGDR, while the differences in the coupling strength between one particle-one hole (1p-1h) and two particle-two hole (2p-2h) correlated (relativistic) and non-correlated (non-relativistic) configurations lead to very different pictures of the importance of the spreading width resulting in wavelet scales being a sensitive measure of their interplay.
The relativistic approach with particle-vibration coupling, in particular, shows impressive agreement with the number and absolute values of the scales extracted from the experimental data.
 
\end{abstract}


\end{frontmatter}

\section{Introduction}
\label{sec:intro}

Giant resonances are elementary excitations of the nucleus and their understanding forms a cornerstone of microscopic nuclear theory.
They can be classified according to their quantum numbers (angular momentum, parity and isospin). 
Gross properties like energy centroids and strengths in terms of exhaustion of sum rules have been investigated extensively in inelastic scattering experiments with electromagnetic and hadronic probes and found to be fairly well described by microscopic models \cite{har01}.
However, a systematic understanding of the decay widths is still lacking. 

The giant resonance width $\Gamma$ is determined by the interplay of different mechanisms: fragmentation of the elementary 1p-1h excitations (called Landau fragmentation or damping $\Delta E$), direct particle decay out of the continuum (escape width $\Gamma\!\uparrow$), and statistical particle decay due to coupling to 2p-2h and many particle-many hole (np-nh) states (spreading width  $\Gamma\!\downarrow$) such that
\begin{equation}
\label{eq:width}
\Gamma = \Delta E + \Gamma\!\uparrow + \Gamma\!\downarrow.
\end{equation}
The coupling of 1p-1h excitations with more complex states implies a fragmentation of the giant resonance strength in a hierarchical manner with characteristic life times, respectively energy scales \cite{bbb98}.
Such a scheme underlies all transport models of quantum many-body systems with applications e.g.\ in nuclear physics, cosmology or condensed matter physics \cite{cas21}.
This ``doorway state'' picture has been widely used in nuclear physics but experimental evidence is scarce. 

One possible option to gain insight into the role of the different components are experiments where the particle decay out of the giant resonance is measured in coincidence. 
Direct decay can be identified by the preferential population of 1h and 1p-2h states in the daughter nucleus, and the spreading width contribution can be estimated by comparison with statistical model calculations (see, e.g.,  Refs.~\cite{bol88,die94,str00,car01,hun03}).
In recent years, an alternative method has been developed based on a quantitative analysis of the fine structure of giant resonances observed in high energy-resolution experiments \cite{vnc19a}.
Different approaches for an extraction of energy scales characterizing the fine structure phenomenon have been compared in Ref.~\cite{she08} and wavelet analysis has been identified as particularly promising.
However, any interpretation of such characteristic energy scales requires comparison to model calculations incorporating some or all of the aforementioned mechanisms.

Fine structure of giant resonances has been established as a general phenomenon for all types of resonances \cite{vnc19a} and studied extensively across the nuclear chart for the cases of the IVGDR \cite{pol14,fea18,jin18,bas20,don20} and the IsoScalar Giant Quadrupole Resonance (ISGQR) \cite{she04,she09,usm11a,usm16,kur18}. 
The source of fine structure was found to be quite different for these two cases. 
Fine structure of the ISGQR arises from the spreading width by coupling of the 1p-1h states to low-lying collective phonons, while for the IVGDR it is dominated by the fragmentation of the 1p-1h strength, i.e, Landau damping \cite{vnc19a}.
Some signatures of the coupling between 1p-1h and 2p-2h states were also found. 

The present work aims at an in-depth study of the role of Landau damping versus spreading width based on wavelet analysis of high energy-resolution photo-absorption data extracted from relativistic Coulomb excitation in the (p,p$^\prime$) reaction at very forward scattering angles \cite{vnc19b}.
We focus on experimental data for $^{40}$Ca and $^{48}$Ca, since doubly-magic nuclei are particularly suited for models including degrees of freedom beyond the mean-field approximation of Random-Phase Approximation (RPA). 
Specifically, we test three approaches allowing for the inclusion of 2p-2h states, viz.\ a realistic interaction derived with the Unitary Correlation Operator Method (UCOM) from the Argonne V18 potential \cite{pap09,pap10}, the time blocking approximation developed for relativistic energy density functionals (RTBA) \cite{lit08} and the {\it ab initio} In-Medium Similarity Renormalization Group (IM-SRG) in combination with RPA and Second RPA (SRPA) \cite{dro90} using two- plus three-nucleon interactions from chiral Effective Field Theory (EFT) \cite{tri16}.

\section{Experiment and determination of equivalent photo-absorption spectra}
\label{sec:exp-results}

High energy-resolution studies of the IVGDR can be performed with (p,p$^\prime$) scattering at energies of several hundred MeV and at extreme forward scattering angles, where relativistic Coulomb excitation dominates the cross sections \cite{vnc19b}. 
Special facilities permitting the combination of a magnetic spectrometer placed at $0^\circ$, detecting the scattered protons, and dispersion-matched beam have been developed \cite{tam09,nev11}.  

\begin{figure}[t]
\begin{center}
	\includegraphics[width=\columnwidth]{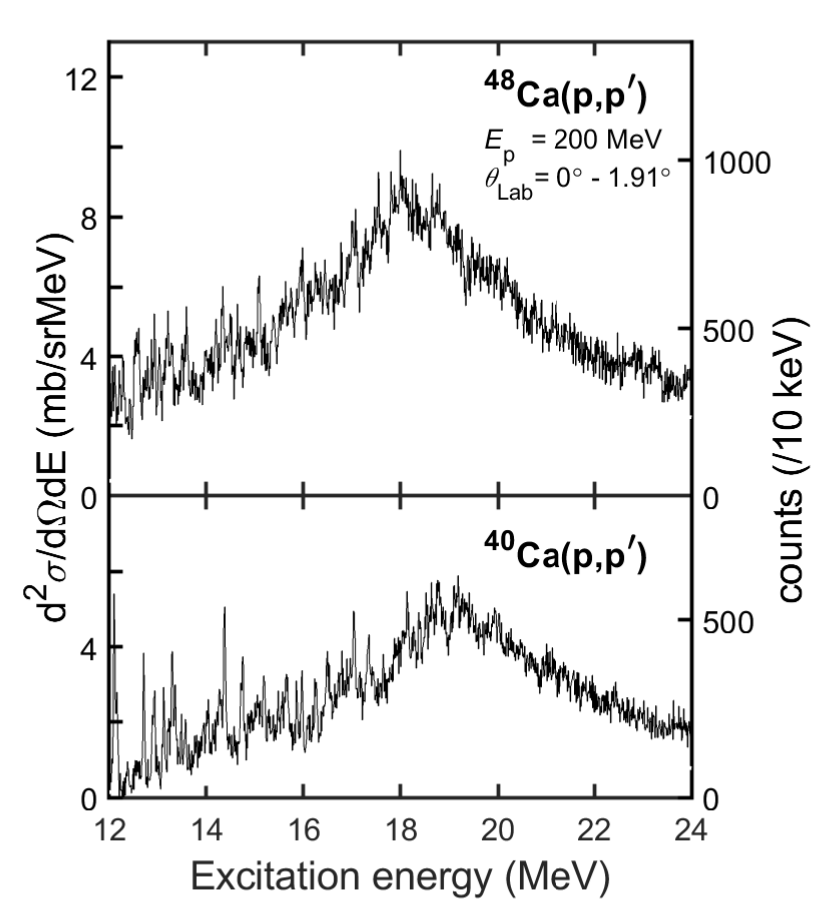}
	\caption{Spectra of $^{40,48}$Ca(p,p$^\prime$) scattering at $E_{\rm p}$ = 200 MeV and scattering angles $\theta_{\rm Lab} = 0^{\circ} - 1.91^{\circ}$. \label{fig:Fig_1}
	}
	\label{fig:Spectrum}
\end{center}

\end{figure}

The present experiments were performed using a dispersion-matched 200 MeV proton beam produced by the Separated Sector Cyclotron (SSC) at the iThemba Laboratory for Accelerator Based Sciences (iThemba LABS), Cape Town, South Africa. 
The data for $^{40}$Ca shown in the bottom part of Fig.~\ref{fig:Spectrum} were obtained during the experiment described in Ref.~\cite{jin18}, but they have been reanalyzed to improve background subtraction and energy resolution.  
For the $^{48}$Ca data, protons were inelastically scattered off a self-supporting target (isotopically enriched to 90.0\%) with an areal density of 1.5 mg/cm$^{2}$. 
Reaction products were momentum-analyzed by the K600 magnetic spectrometer in $0^\circ$ mode \cite{nev11} with the acceptance defined by a circular collimator with an opening angle of $\theta_{\mathrm{Lab}}=0^\circ \pm 1.91^{\circ}$. 
Further details of the $^{48}$Ca  experiment and data extraction are given in Ref.~\cite{lat17}.
The resulting spectrum shown in the top part of  Fig.~\ref{fig:Spectrum} demonstrates pronounced fine structure up to excitation energies of about 20 MeV as also seen in $^{40}$Ca. 
The energy resolution achieved is $\Delta E \simeq 40$ keV (Full Width at Half Maximum (FWHM)) in both spectra.
The scale on the right side shows the counting statistics which 
reach several hundred per 10 keV bin, sufficient for an analysis of the fine structure.

The spectra were converted to equivalent photo-absorption cross sections with the methods described in Refs.~\cite{don18,don20,vnc19b} to facilitate direct comparison with theoretical predictions.
This involves nuclear background subtraction and  determination of the virtual-photon production function for the conversion from Coulomb to photo-absorption cross sections. 
Besides Coulomb cross sections, the spectra contain contributions from the IsoScalar Giant Monopole Resonance (ISGMR) and the ISGQR lying under the IVGDR. 
In addition, there is a contribution from quasi-free scattering (see e.g.\ Ref.~\cite{bir17}) which increases towards the higher excitation-energy end of the IVGDR. 
Possible contributions of the ISGMR and ISGQR to the measured cross sections were estimated following the method outlined in Ref.~\cite{don18} using the experimental strength distributions of Ref.~\cite{how20} and found to be well below 10\% at their respective maxima.
In the study of the IVGDR fine structure in $^{208}$Pb \cite{pol14}, it was demonstrated that a discrete wavelet analysis \cite{she08,kal06} provides a good approximation for the nuclear background to be subtracted and the same type of analysis was performed in the present case.  

The conversion of Coulomb to photo-absorption cross sections was based on the virtual photon method. 
The virtual $E1$ photon spectra were calculated within the Eikonal approximation \cite{ber93} and the resulting photo-absorption spectra are in good agreement with previous work (Ref.~\cite{ahr75} for $^{40}$Ca and Refs.~\cite{str00,bir17} for $^{48}$Ca).

\section{Theoretical models}
\label{sec:theory}

In the following, we discuss three models for a comparison of $E1$ strength functions with the experiment results.

\subsection{Relativistic approaches with an effective meson-exchange interaction}
\label{theory:pp}

Relativistic RPA (RRPA) was formulated, implemented and further investigated in Refs.~\cite{Vretenar1999,Ma2001} as a dynamical extension of relativistic mean-field theory. The time-dependent relativistic Hartree approximation was obtained for the Dirac spinors self-consistently with the classical meson fields that lead to the RRPA in the small-amplitude limit. The RRPA, as well as its extension to superfluid  systems \cite{PaarRingNiksicEtAl2003}, successfully describe the positions of nuclear collective modes, above all those of giant resonances, in the no-sea approximation, which imply transitions between nucleonic states of the Fermi and Dirac sectors of the energy domain. The sum rules are also described fairly well.

Major upgrades were proposed in Refs.~\cite{LitvinovaRingTselyaev2007,LitvinovaRingTselyaev2008,LitvinovaRingTselyaev2010}, where the RRPA  was extended, for the first time, by the (quasi)Particle-Vibration Coupling (qPVC) mechanism in a fully self-consistent, i.e., parameter-free scheme. The approach was first derived with the aid of the time blocking technique and named Relativistic (Quasiparticle) Time Blocking Approximation (R(Q)TBA).
Later, the R(Q)TBA was re-derived in the model-independent Equation Of Motion (EOM) framework based on the bare fermionic Hamiltonian without applying time blocking operators \cite{LitvinovaSchuck2019}. The EOM formalism allowed for an {\it ab initio} description and for extended approaches beyond the leading qPVC dynamical kernels. Both the original and extended versions of the R(Q)TBA demonstrated significant improvements in the description of nuclear collective excitations. Most remarkably, the qPVC already provides a reasonable degree of fragmentation of the 2-quasiparticle states in the leading approximation \cite{LitvinovaRingTselyaev2007,LitvinovaRingTselyaev2008}. Moreover, the description of the low-energy (soft) modes was refined considerably
\cite{LitvinovaLoensLangankeEtAl2009,EndresLitvinovaSavranEtAl2010,EndresSavranButlerEtAl2012,LanzaVitturiLitvinovaEtAl2014,Oezel-TashenovEndersLenskeEtAl2014,EgorovaLitvinova2016}.

In this work, we apply the original version of the RTBA \cite{LitvinovaRingTselyaev2007} without pairing correlations since the calcium isotopes considered here are of doubly-magic nature. 
The numerical implementation is grounded in the self-consistent Relativistic Mean Field (RMF) with NL3 forces \cite{Lalazissis1997}, which determines the meson and nucleon fields, and the single-nucleon Dirac-Hartree basis. In this basis, the RRPA equations were solved to obtain the vertices and frequencies of the phonon modes. The phonon model space was truncated using the same criteria as in the series of earlier calculations \cite{EgorovaLitvinova2016}.  The selected phonons, together with the single-particle output of the RMF, determine the PVC amplitude of the RTBA and, further, the strength function in the $J^{\pi} = 1^-$ channel. 

\subsection{In-Medium RPA and SRPA with ab initio interactions}
\label{theory:rr}

As a second class of methods for the calculation of the dipole response, we employ RPA and SRPA in conjunction with the In-Medium Similarity Renormalization Group (IM-SRG) using realistic two plus three-nucleon interactions from chiral EFT. 

The IM-SRG is initially an {\it ab initio} method to obtain the ground-state energies of closed-shell nuclei through a unitary decoupling of a Slater-determinant reference state from particle-hole excitations. The decoupling is implemented via a similarity renormalization-group flow equation for the matrix elements of the Hamiltonian, normal-ordered with respect to the reference state. As a function of a continuous flow-parameter, the matrix elements that connect the reference state to particle-hole excitations are successively suppressed, i.e., the Hamiltonian is partially diagonalized in a particle-hole basis. Since we use a unitary transformation, the eigenvalues of the Hamiltonian are preserved and despite the simplicity of the ground or reference state, which remains a Slater determinant, all correlation effects are absorbed into the transformed Hamiltonian. The IM-SRG has become a standard tool for {\it ab initio} calculations of ground states for closed-shell nuclei in the medium-mass regime \cite{tsu10,her12,her15}. 

For the description of collective excitations, we supplement the IM-SRG with a second many-body method that provides access to excited states. 
Here, we use RPA and SRPA on top of the IM-SRG reference state. The input for the (S)RPA calculations is the IM-SRG evolved Hamiltonian with complete ground-state decoupling, which has profound consequences. Since the IM-SRG suppresses matrix elements connecting the ground state to 1p-1h and 2p-2h excitations, the backward amplitudes in (S)RPA are also suppressed, effectively reducing the (S)RPA to the second Tamm-Dancoff approximation. This is not surprising, since ground-state correlations have been absorbed by the IM-SRG transformation. In SRPA, the suppression of the coupling to 2p-2h states, which goes beyond the Brillouin condition of Hartree-Fock (HF), implies the absence of instabilities \cite{pap14,tse13}. We refer to these hybrid methods as IMRPA and IMSRPA, where more details can be found in Ref.~\cite{tri16}. 

For the following calculations, we use the N2LO$_\text{SAT}$ interaction \cite{eks15} with a free space SRG evolution to $\alpha=0.08\,\text{fm}^4$ \cite{rot14}. All calculations are performed in model spaces including 13 harmonic-oscillator shells. 

\subsection{RPA and SRPA with the UCOM interaction}
\label{theory:pp}

The isovector dipole distributions are also calculated using the RPA and SRPA formalisms with the UCOM two-nucleon effective interaction derived from the realistic Argonne V18 interaction \cite{pap09,pap10}. 
The same models were employed in Ref. \cite{usm11a} to analyze the fine structure of the ISGQR in $^{40}$Ca. 
UCOM-SRPA generally provides a good description of giant dipole and quadrupole resonances in closed-shell nuclei \cite{pap09}. Here, we use a single-particle basis of 13 harmonic-oscillator shells to solve the HF equations for the reference state and to obtain the HF single-particle basis.
Then, all possible 1p-1h and 2p-2h configurations that can be built from the HF basis states and coupled to the $1^-$ quantum numbers are used to construct the RPA and SRPA spaces and to solve the respective eigenvalue problems. 
Couplings between 1p-1h and 2p-2h configurations are fully included in SRPA as well as p-p and h-h interactions in the 2p-2h propagator. 
The SRPA calculations can thus describe an incoherent damping mechanism from 2p-2h coupling, in addition to Landau damping.

\section{Wavelet analysis technique}
\label{sec:wavelet}

Extensive details of the wavelet analysis technique using the Continuous Wavelet Transform (CWT) as applied to the analysis of fine structure in high energy-resolution giant resonance investigations can be found in Refs.~\cite{she08,jin18,don20}. 
As such, only brief details are given here. 
The response of the K600 focal-plane detector is approximated well by a Gaussian lineshape. 
Typically, for the analysis of giant-resonance fine structure from the measured excitation-energy spectra, the Morlet mother wavelet, being a Gaussian envelope on top of a periodic structure, is the most suitable. 
In the present fine-structure analysis, the Complex Morlet wavelet was used (see Ref.~\cite{coo08}, Fig. 1)
\begin{equation}
\label{eq:complexMor}
\Psi(x) = \frac{1}{\sqrt{{\pi}f_{\mathrm{b}}}} \exp(2{\pi}i{f_{\mathrm{c}}}x) \exp \left( - \frac{x^2}{f_{\mathrm{b}}} \right).
\end{equation}
Here, $f_{\mathrm{b}}$ controls the wavelet bandwidth and $f_{\mathrm{c}}$ the center frequency of the wavelet.

%
\begin{figure}[t]
\begin{center}
	\includegraphics[width=\columnwidth]{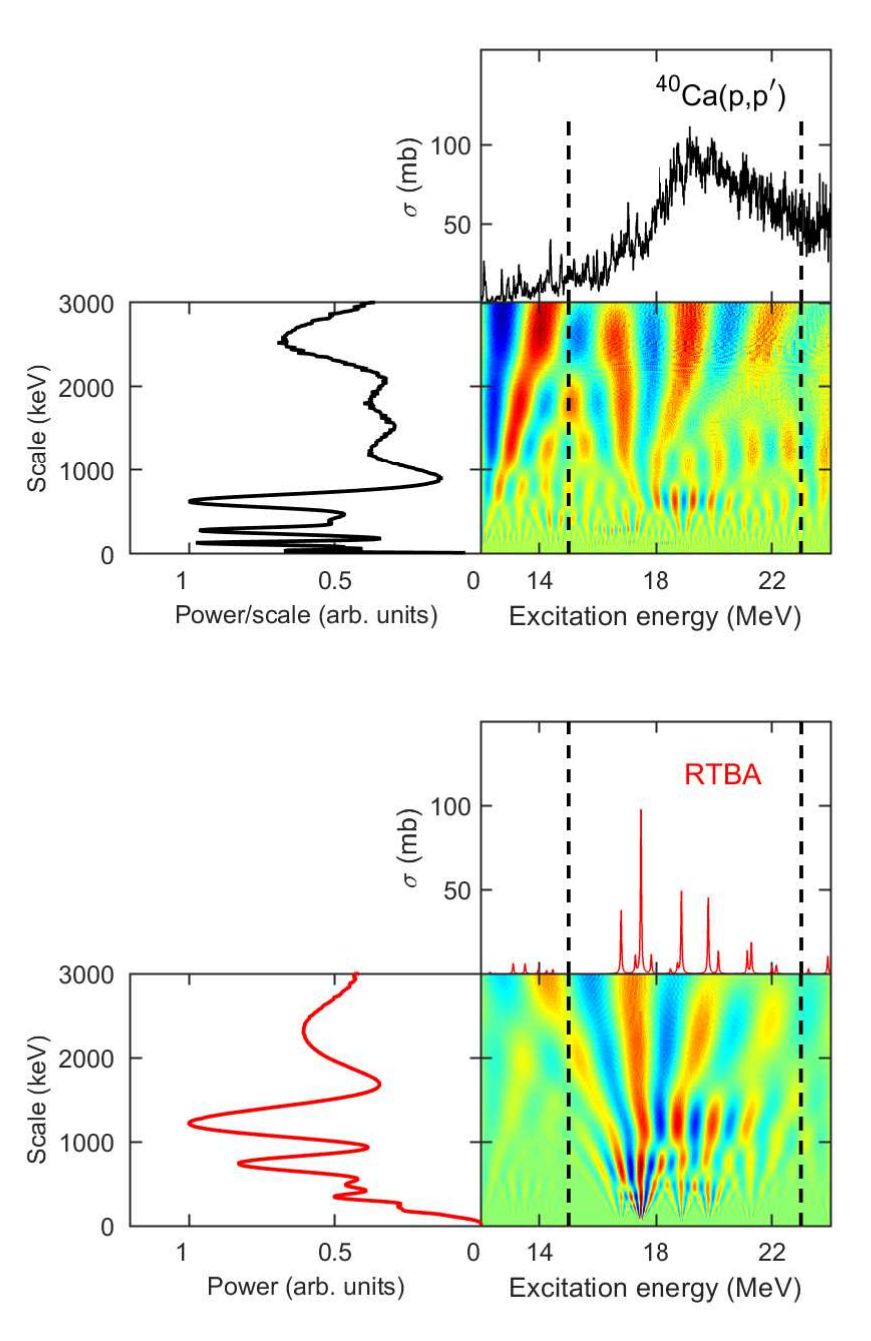}
	\caption{Wavelet analysis of the $^{40}$Ca(p,p$'$) equivalent photo-absorption cross sections (top) and  of the $^{40}$Ca RTBA $E1$ strength function (bottom). 
	For details see text.
	\label{fig:Fig_2}
	}
	\label{fig:scales}
\end{center}
\end{figure}
\begin{figure*}
\begin{center}
  \includegraphics[width=15cm]{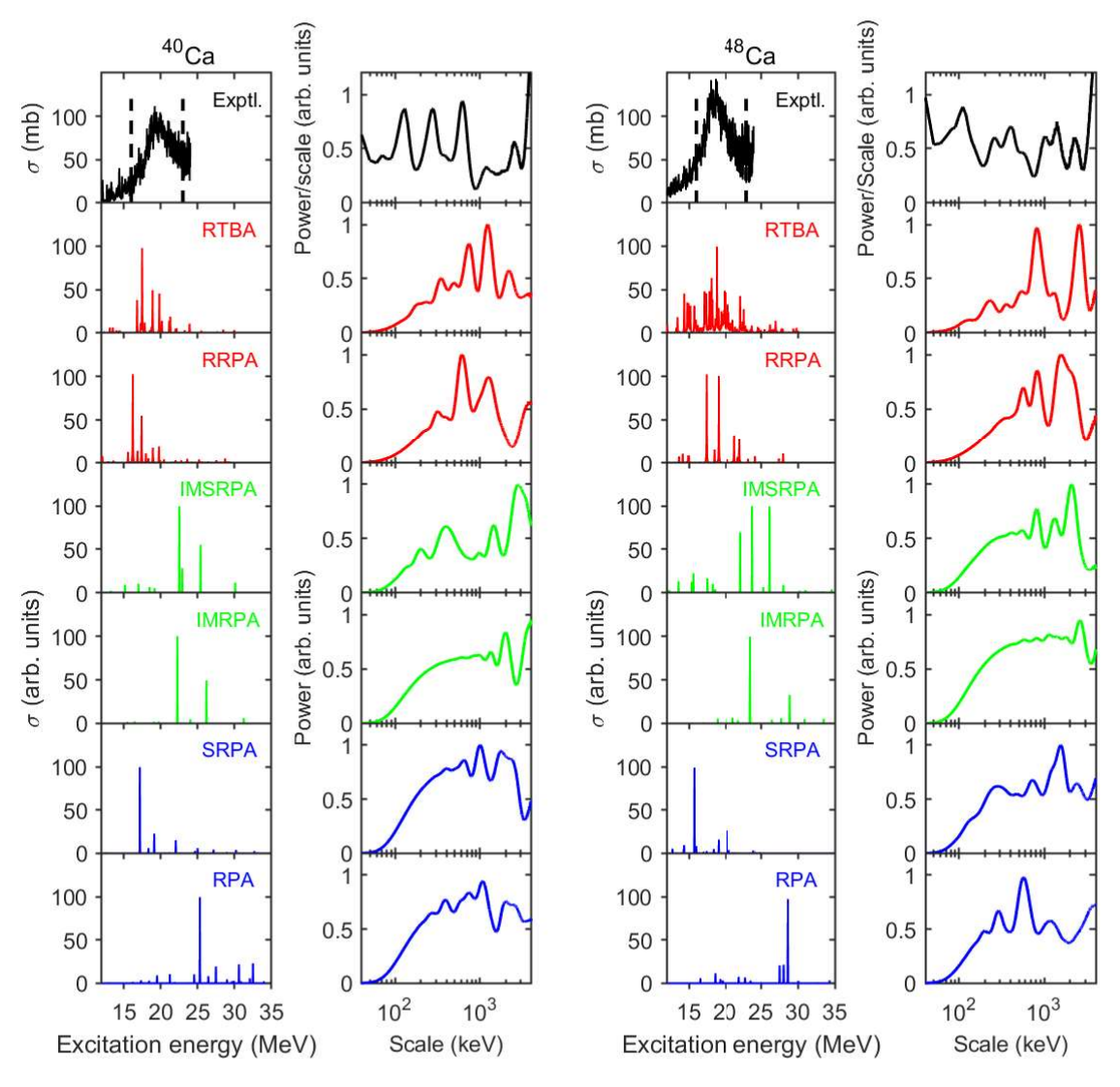}
	\caption{Wavelet analysis for $^{40,48}$Ca equivalent photo-absorption cross sections  and corresponding $\it{E}$1 strength functions. 
    }
\label{fig:ExpTheoAll}
\end{center}
\end{figure*}

By way of example, the real parts of the complex wavelet coefficients determined for the equivalent photo-absorption spectrum for $^{40}$Ca (Fig.~\ref{fig:scales} top right-side upper panel) are shown just below in a two-dimensional plot as a function of excitation energy and scale.
Large positive values are indicated by shades of red going down to close to zero in yellow with negative coefficients in shades of blue.

Wavelet energy scales can be extracted from the wavelet coefficient plot as peaks in the corresponding power spectrum obtained by squaring and summing the complex CWT coefficients
\begin{equation}
P({\delta}E) = {\frac{1}{N}}\sum_i \vert C_i(\delta E) C_i^*(\delta E) \vert \;\, ,
\label{eq:power}
\end{equation} 
where $P({\delta}E)$ is the power as a function of scale ${\delta}E$ summed at each scale value over the index $i = N$ with $N$ being the number of energy bins on the excitation-energy axis. 
In the present case, the range of summation is shown by lower and upper excitation-energy limits indicated by the vertical dashed lines in Fig.~\ref{fig:scales}, top right-side panels. 
The corresponding power spectrum obtained from Eq.~(\ref{eq:power}) is shown in Fig.~\ref{fig:scales}, top left-side panel. Here, for clarity, power $P({\delta}E)$ is divided by the corresponding scale ${\delta}E$ and the result is normalized to unity. 
Continuing the CWT analysis example, the procedure above was repeated for the $^{40}$Ca RTBA calculations, the results of which are displayed in the bottom section of Fig.~\ref{fig:scales}. 

With respect to the scale values extracted and discussed below, we note that these do not depend on a particular choice of the wavelet function (see Fig.~9 in Ref.~\cite{she08} for an example) but the complex-Morlet function provides the best compromise between resolution in excitation energy and scale for the kind of data analyzed here.
Furthermore, it yields the equivalent Fourier scale. 
A particular value of scale in a CWT plot corresponds to the excitation-energy difference between consecutive minima (or maxima) in the coefficient plot, referred to as a ``length-like'' wavelet energy-scale. 
Half of the ``length-like'' wavelet energy-scale is the width of the peak (FWHM) and is referred to as a ``width-like'' wavelet energy-scale. 
We also note that the model dependence of the background subtraction with the DWT (Sec.~\ref{sec:exp-results}) discussed e.g.\ in Ref.~\cite{usm11a} does not affect the scale values but only their relative magnitude.

\section{Damping of the IVGDR - wavelet energy-scales comparison}
\label{sec:results}

The experimental equivalent photo-absorption cross sections and model $E1$ strength functions together with corresponding wavelet analysis power-spectra are shown in Fig.~\ref{fig:ExpTheoAll} for $^{40}$Ca (left-side two columns) and $^{48}$Ca (right-side two columns). 
Lower and upper excitation energy limits for determination of the power spectra are shown by the vertical dashed lines in the $^{40,48}$Ca equivalent photo-absorption spectra.
These same limits are used for the corresponding $E1$ strength functions RTBA and RRPA (red), and SRPA (blue). However, since the $\it{ab~initio}$ IMSRPA and IMRPA (green), and the RPA (blue) $E1$ strengths are shifted up in excitation energy, the full excitation-energy range shown was used to calculate the corresponding power spectra. As can be seen, many scales can be identified from the peaks present in the various power spectra. 

Comparison of the experimental and theoretical photo-absorption cross sections shows that the calculations based on relativistic Lagrangians describe the IVGDR centroids well, in particular when 2p-2h degrees of freedom are included.
The results based on chiral interactions are systematically too high by 3-4 MeV for IMSRPA and IMRPA(green) and significantly too high at the RPA(blue) level for the realisitic interaction, while the SRPA(blue) results are downshifted slightly below the experimental peak of the IVGDR.
The effect of coupling to 2p-2h degrees of freedom is strong for RTBA(red), especially for $^{48}$Ca, sizable for IMSRPA(green) and weak for the SRPA(blue) calculations based on the realistic interaction.
On the other hand, the effect is also stronger in $^{48}$Ca than in $^{40}$Ca in UCOM-SRPA. 
Inspection of the configuration content of the eigenstates of the latter results reveals that 1p-1h configurations account for $90\%$ of the norm in $^{40}$Ca in this energy region, while in $^{48}$Ca many eigenstates are made of mostly 2p-2h configurations. 
Those states do not contribute significantly to the photo-absorption cross section, which is calculated here by applying the usual single-particle dipole excitation operator. 

All calculations exhibit characteristic scales already on the RPA level, which is consistent with findings in lighter \cite{jin18,fea18} and heavier \cite{pol14,bas20} nuclei.
While the scale region around 1 MeV is not so much affected, inclusion of 2p-2h states in the RTBA(red) and IMSRPA(green) calculations leads to the appearance of new peaks in the power spectra at lower scale values.
The  UCOM-SRPA results, on the other hand, show mainly a  redistribution of scales with respect to the RPA calculations and no scales at values below 400 keV. 

%
\begin{figure}[t]
\begin{center}
	\includegraphics[width=\columnwidth]{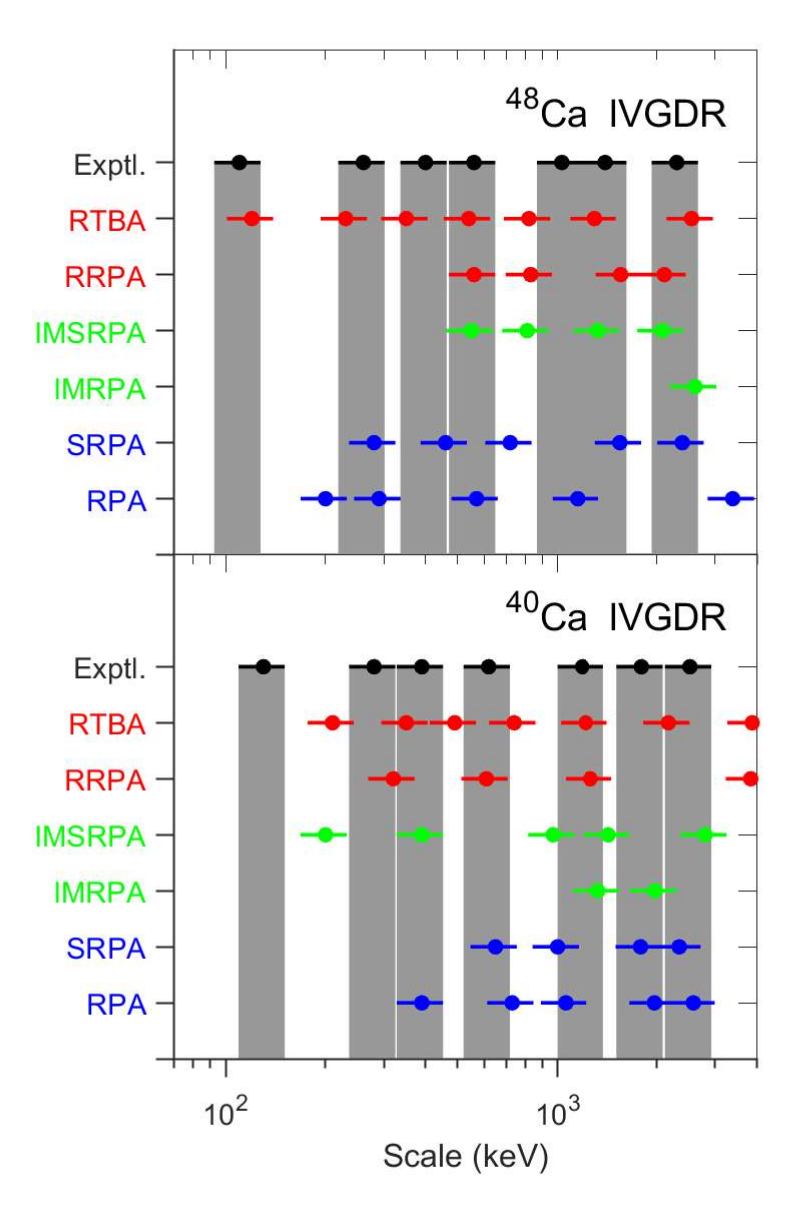}
	\caption{Wavelet-analysis energy scales comparison for $^{40,48}$Ca(p,p$'$) equivalent photo-absorption cross sections and corresponding $\it{E}$1 strength functions. 
	}
\label{fig:ScalesComp}
\end{center}
\end{figure}

In order to facilitate a quantitative comparison between experimentally determined scales and the various model predictions, Fig.~\ref{fig:ScalesComp} shows the position of extracted experimental scales (power spectrum peaks, ``length-like'' scale) as filled black circles together with an error bar (corresponding ``width-like'' scale converted to standard deviation). For the theoretical comparisons, these experimental scales are then indicated by the grey areas 
%
on top of which are plotted the extracted scales for the various model predictions following the same colour code as in Fig.~\ref{fig:ExpTheoAll}. 
Note that, for the sake of clarity since many scales from experiment and theory have been identified, the total width of the scale displayed has been reduced to 1.5 standard deviations. 

Both experimental results show a similar picture in the number of scales and their distribution in energy.
We have tested whether the scales are independent of excitation energy by repeating the analysis with gates on the low-energy (16-20 MeV) and high-energy (20-24 MeV) regions of the main IVGDR peaks. 
The scale energies are found to be approximately the same, but the relative power changes.
The model predictions reveal large differences between $^{40}$Ca and $^{48}$Ca.
In $^{48}$Ca, all models find a significant effect of the spreading width when going from 1p-1h to 1p-1h plus 2p-2h calculations by producing new scales at values $<$1 MeV. 
The RTBA result is particularly impressive by not only reproducing the correct number of scales but also their magnitude within the assigned uncertainty.
This also includes a scale at very small values ($<$200 keV) which was found to be a generic signature of the spreading width observed in all cases experimentally studied so far, independent of the type of resonance or the mass region \cite{vnc19a}. 

\section{Conclusions}
\label{sec:conclusion}

We use characteristic energy scales derived from a wavelet analysis of high energy-resolution photo-absorption data to investigate the role of Landau fragmentation and spreading width in the damping of the IVGDR.
The doubly-magic nuclei $^{40,48}$Ca were chosen to minimize the influence of ground-state correlations.
The experimental results are compared to three state-of-the-art models allowing for the inclusion of 2p-2h degrees of freedom.
These are based on a relativistic approach with an effective meson-exchange interaction, SRPA with the UCOM effective interaction and, for the first time, on chiral EFT including three-nucleon interactions.
The models show significant differences in the coupling of 1p-1h and 2p-2h states and correspondingly the importance of the spreading width for an understanding of wavelet scales.
A remarkable agreement for the number of scales and their absolute values is achieved with the RTBA calculations, in particular for $^{48}$Ca.
Although the calculations based on IM-SRG and a chiral EFT interaction still show some deficiencies in reproducing gross features, they are very encouraging for the development of {\it ab initio}-based models beyond RPA.
The results demonstrate that high energy-resolution data combined with a wavelet analysis can provide unique insight into the role of Landau fragmentation and spreading width in the damping of the IVGDR.

\section*{Acknowledgements}

This work was supported by the South African National Research Foundation (NRF) through  Grant Nr.~85509 and by the Deutsche Forschungsgemeinschaft (DFG, German Research Foundation) under Grant No. SFB 1245 (project ID 279384907).
E.L. acknowledges financial support from the National Science Foundation of the United States of America US-NSF under the CAREER Award PHY-1654379.
The work of P.~Papakonstantinou was supported by the Rare Isotope Science Project of the Institute for Basic Science funded by the Ministry of Science, ICT and Future Planning and the National Research Foundation (NRF) of Korea (2013M7A1A1075764).



\end{document}